\documentclass{lmcs} 
\pdfoutput=1

\usepackage{lastpage}
\lmcsdoi{22}{2}{12}
\lmcsheading{}{\pageref{LastPage}}{}{}%
{Jun.~27,~2025}{May~01,~2026}{}


\usepackage[utf8]{inputenc}

\usepackage{amsmath}
\usepackage{amssymb}
\usepackage{amsthm} 	
\usepackage{amsfonts}
\usepackage{mathrsfs}

\usepackage{enumerate}

\usepackage{microtype}  
\usepackage{url}

\usepackage{graphicx}
\usepackage{xcolor}

\newcommand{\nat}[1]{[#1]}          

\DeclareMathOperator{\pred}{\mathrm pred}

\newcommand{\NS}{\ensuremath{\mathrm{NS}}}
\newcommand{\MC}{\ensuremath{\mathrm{MC}}}
\newcommand{\PC}{\ensuremath{\mathrm{PC}}}

\usepackage{hyperref}   
\usepackage{bm}   

\usepackage{nicefrac}

\newcommand{\peb}{\mathrm {Peb}}
\newcommand{\var}{\mathrm {Var}}

\newcommand{\twopartdefotherwise}[4] 
{
    \left\{
    \begin{array}{ll}
        #1, & \mbox{if } #2 \\
        #3, & \mbox{otherwise} #4
    \end{array}
    \right.
}

\newcommand{\VS}{\mathrm {VSpace}}
\newcommand{\Size}{\mathrm {Size}}

\usepackage{bussproofs} 
\usepackage[boxempty]{stmaryrd}

\newcommand{\gamevalueformat}[1]{\mathsf{#1}}

\newcommand{\Pebbling}{\mathcal{P}}
\newcommand{\PebConfig}{\mathbb{P}}

\newcommand{\pebspace}{\gamevalueformat{space}}
\newcommand{\pebtime}{\gamevalueformat{time}}

\newcommand{\BW}{\gamevalueformat{BW}}
\newcommand{\Black}{\gamevalueformat{Black}}
\newcommand{\White}{\gamevalueformat{White}}
\newcommand{\Rev}{\gamevalueformat{Rev}}
\newcommand{\Res}{\gamevalueformat{Res}}
\newcommand{\pebtimestep}{i} 
\newcommand{\pebendtime}{t} 

\begin{document}

\title{Pebble Games and Algebraic Proof Systems}

\author[L.M.~Jaser]{Lisa-Marie Jaser}[a]
\author[J.~Tor\'an]{Jacobo Tor\'an}[a]

\address{Universit\"at Ulm, Germany}
\email{lisa-marie.jaser@uni-ulm.de, jacobo.toran@uni-ulm.de}






\begin{abstract}
    Analyzing refutations of  the well known
    pebbling formulas $\peb(G)$    we prove some new strong connections between pebble games and algebraic proof system,  showing that
    there is a parallelism between the reversible, black and black-white pebbling games on one side, and
    the three algebraic proof systems Nullstellensatz, Monomial Calculus and Polynomial Calculus  on the other side.  In particular we prove that for any directed acyclic graph $G$ with a single sink, if there is a Monomial Calculus refutation 
for  $\peb(G)$ having simultaneously degree $s$ and size $t$ 
then there is a black pebbling strategy on $G$ with space $s$ and time $t+s$.
Also if there is a black pebbling strategy for $G$ with space $s$ and time $t$ it is possible to extract from it
a MC refutation 
for  $\peb(G)$ having simultaneously degree $s$ and size $ts$. 
These results are analogous to those proven in \cite{RezendeMNR21} for the case of reversible pebbling and
Nullstellensatz.
Using them  we  prove degree separations between Nullstellensatz, Monomial Calculus
 and Polynomial Calculus, as well as strong  degree-size tradeoffs for Monomial Calculus.

We also 
notice that for any directed acyclic graph $G$ the  space
needed in a pebbling strategy on $G$, for the three versions of the game, reversible, black and black-white, exactly matches the variable space complexity of a refutation of the corresponding pebbling formula $\peb(G)$ in each of the 
algebraic proof systems Nullstellensatz, Monomial Calculus and Polynomial Calculus.
Using known pebbling bounds on graphs, this connection implies
separations between  the corresponding variable space measures. 
\end{abstract}

\maketitle

\section{Introduction}
The use of pebble games 
in complexity theory goes back many decades. They offer 
a very clean tool to analyze certain complexity measures, mainly space and time, in an isolated way 
on a graph, which can then be translated to specific computational models. 
Very good overviews of  these results can be found in 
\cite{Pippengder80PebblingWITHOUTNOTE,Savage98,Nordstrom09PebblingSurvey}. 

We consider several versions of the game,
defined formally  in the preliminaries.
Intuitively, the goal  of these games is to measure the minimum  number of pebbles needed by a single player in order to place a pebble on the  sink of a directed acyclic graph (DAG) following certain rules (this is called the pebbling price).
A black pebble can only be placed on a vertex if it is a source or if all its direct predecessors already have a pebble on them, 
but these pebbles can be removed at any time.
A white pebble (modelling non-determinism) can be placed on any vertex $v$ at any time but can only be removed if all its direct predecessors contain a pebble, which is considered as a certificate for  the fact that the pebble could have been placed on $v$.
In the reversible pebble game (modelling reversible computation), pebbles can only be placed or removed from a vertex if all the direct predecessors of the vertex contain a pebble.
These three games define a short hierarchy being reversible pebbling weaker than
black pebbling and this in turn weaker than the black-white pebble game. 

In proof complexity one tries to understand the resources needed
for a proof of a mathematical statement in a formalized system.
Pebbling games have also become one of the most useful tools for proving results in this area, as can be seen for example in
\cite{Nordstrom06NarrowProofsMayBeSpaciousSTOC},\cite{BN11UnderstandingSpaceFULLREF}, \cite{Nordstrom12RelativeStrength}, \cite{Nordstrom13SurveyLMCS}, \cite{ToranW21}, \cite{RezendeMNR21}, to name just a few references.
The reason for this is that one can often translate a certain 
measure for the pebbling game, mainly number of pebbles or pebbling time, 
into a suitable complexity measure for a concrete proof system. Very often
the bounds for this measure in a graph translate accurately to complexity bounds in the different proof systems for 
a certain kind of contradictory formulas mimicking the game, called pebbling formulas. These formulas were introduced in~\cite{BW01ShortProofs} and have
been extremely useful for proving separations, upper and lower bounds as well as tradeoff results in basically all studied proof systems.  
See e.g. \cite{Nordstrom13SurveyLMCS}.  

In the present paper we will concentrate on algebraic proof system. In these
systems   formulas are encoded  as  sets of 
polynomials over a field and the question of whether 
 a formula is unsatisfiable is translated to  the question 
 of whether the polynomials have a common root. Powerful algebraic tools
 like the Gr\"obner Basis Algorithm can be used for this purpose. 
 Several algebraic proof systems have been introduced in the literature (defined
 formally bellow). Well known ones are Nullstellensatz  (NS)  introduced in 
 \cite{BIKPP94LowerBounds} and the more powerful Polynomial Calculus (PC)
 defined in \cite{CEI96Groebner}. The first one is usually considered as a static system in which a ``one-shot" proof has to be produced, while in PC
 there are certain derivation rules like in a more standard proof system.

 The best studied  complexity measures for refutations in 
 these systems are the degree (maximum
 degree of a polynomial in the refutation) and size (number of monomials in the refutation, counted with repetitions).
 For studying the connections with the pebble games it is very useful to consider also  space measures and the configurational
 refutations associated with space. We will use the variable space 
 measure (number of variables that appear simultaneously in a refutation). 
 
 In \cite{BerkholzG15} the Monomial Calculus system (MC) was identified. This system is defined by limiting the multiplication rule in PC to monomials, and
 its power lies between $\NS$ and $\PC$. Building on results from \cite{AM13SheraliAdamsRelaxations} for the Sherali-Adams proof system,
 the authors proved that for  any pair
 of non-isomorphic graphs, the MC degree for the refutation of the corresponding isomorphism formulas 
exactly  corresponds to
the Weisfeiler-Leman bound for separating the graphs, a very important tool
in graph theory and descriptive complexity. 
This equivalence (as well as the relations to pebbling shown here)
motivates the study of  Monomial Calculus as a natural proof system
between NC and PC.

As mentioned above, connections between pebbling games and algebraic systems have been known since  \cite{BCIP02Homogenization}.
It was proved in the mentioned reference that for any directed acyclic graph (DAG)
$G$ the corresponding pebbling formula $\peb(G)$ can be refuted with constant
degree in PC but it requires degree $\Omega(s)$ in NS, where $s$ is the 
black pebbling price of $G, \Black(G)$. Using pebbling results, 
this automatically proves a strong degree separation between NS and PC.
As a more  recent example, the authors in \cite{RezendeMNR21} proved a very tight connection 
between NS and the reversible pebbling game. They  showed that 
space and time in the game played on a DAG exactly correspond to the degree and
size measures in a NS refutation of the corresponding pebbling formula. 
From this connection strong degree-size 
tradeoffs for NS follow. This result also improves the degree separation
from \cite{BCIP02Homogenization} since it is known that there are
graphs for which the reversible pebbling price is a logarithmic factor
larger than the black pebbling price.

We show   in this paper that besides these results, there are further  parallelisms between the reversible, black and 
black-white game hierarchy on one side, and the NS, MC and PC proof systems on the other side.

\subsection{Our Results}

In Section~\ref{MC}
we prove that very similar  results to those given in \cite{RezendeMNR21} for NS and reversible pebbling
are also true for the case of MC and black pebbling. More concretely we show
in Theorem~\ref{thm:MC-ref-peb} that for any DAG $G$ with a single sink, if there is a MC refutation 
for  $\peb(G)$ having simultaneously degree $s$ and size $t$ 
then there is a black pebbling strategy on $G$ with space $s$ and time $t+s$.
This is done by proving that any Horn formula has a very especial kind of MC
refutation, which we call input MC refutation since it is the same concept
as an input refutation in Resolution. Horn formulas constitute 
an important  class with applications in many areas
like  program verification or  logic programming.  It is well known that input 
Resolution is complete for Horn formulas.

For the other direction, we  show in Theorem~\ref{thm:peb-MC} that from a black pebbling strategy for $G$ with space $s$ and time $t$ it is possible to extract
a MC refutation 
for  $\peb(G)$ having simultaneously degree $s$ and size $ts$. 
The small loss in the time parameter compared to the results in \cite{RezendeMNR21} comes from the fact that size complexity is measured in slightly
different ways in NS and MC.
Using these results we are able to show degree separations between NS and MC
as well as the first strong degree separations between MC and PC. 
We also use the simultaneous relation with  time and space in
the black pebbling game to obtain 
strong  degree-size tradeoffs for MC in the same spirit as those in \cite{RezendeMNR21}. 
These results show that there are families of unsatisfiable formulas that have polynomial size MC
refutations with a certain degree, but the size of the refutation becomes superpolynomial
if the degree decreases. 
The results also show that strong degree
lower bounds for MC refutations do not imply exponential size 
lower bounds as happens in the PC proof system \cite{IPS99LowerBounds}.

From the results in \cite{RezendeMNR21} and this paper follows that the degrees of  the refutations of   pebbling formulas in NS and  MS correspond exactly to the space in reversible and black pebbling games respectively.
It would be a nice result if the same could be said about PC degree and space
in the black-white game. Unfortunately this is not the case since as
mentioned above, it was proven in \cite{BCIP02Homogenization} that 
for any DAG the corresponding pebbling formula can be refuted within constant
PC degree. We notice however that if instead of the degree we consider the 
complexity measure of variable space, then  the connection still holds.
We notice that for any single sink DAG $G$
the variable space complexity of refuting $\peb(G)$ in each of the 
algebraic proof systems $\NS, \MC$ and $\PC$ is exactly the
space needed in a strategy for pebbling $G$ in each of the three versions
Reversible, Black and Black-White of the pebble game.
These results allow us to apply known separations between the pebbling space needed in the different versions of the game, in order to obtain the first 
separations in the variable space measure between the different proof systems.

\section{Preliminaries}
\subsection{Pebble Games}
\label{subsec:PebbleGames}

Black pebbling was first mentioned implicitly in~\cite{PH70Comparative}, while black-white pebbling was introduced in~\cite{CS76Storage}. 
Note that there exist several variants of the (black-white) pebble game in the literature.
For differences between these variants, we refer to~\cite{Nordstrom09PebblingSurvey}.
For the following definitions, let $G = (V,E)$ be a DAG with a unique sink vertex $z$.

\begin{defi}[Black and black-white pebble games]
\label{def:classicalBWPebbleGame}
The \emph{black-white pebble game} on $G$ is the following one-player game:
At any time $\pebtimestep$ of the game, there is  a \emph{pebble configuration} $\PebConfig_\pebtimestep := (B_\pebtimestep,W_\pebtimestep)$,
where $B_\pebtimestep \cap W_\pebtimestep = \varnothing$ and $B_\pebtimestep \subseteq V$ is the set of black pebbles and $W_\pebtimestep \subseteq V$ is the set of white pebbles, respectively.
A pebble configuration $\PebConfig_{\pebtimestep-1} = (B_{\pebtimestep-1}, W_{\pebtimestep-1})$ can be changed to $\PebConfig_\pebtimestep = (B_\pebtimestep, W_\pebtimestep)$ by applying exactly one of the following rules:
\begin{description}
    \item[Black pebble placement on $v$:] If all direct predecessors of an empty vertex $v$ have pebbles on them, a black pebble may be placed on $v$. More formally, letting $B_\pebtimestep = B_{\pebtimestep-1} \cup \{ v \}$ and $W_\pebtimestep = W_{\pebtimestep-1}$ is allowed if $v \not \in B_{\pebtimestep-1} \cup W_{\pebtimestep-1}$ and $\pred_G(v) \subseteq B_{\pebtimestep-1} \cup W_{\pebtimestep-1}$. In particular, a black pebble can always be placed on an empty source vertex $s$, since $\pred_G(s) = \varnothing$.
	\item[Black pebble removal from $v$:] A black pebble may be removed from any vertex at any time. Formally, if $v \in B_{\pebtimestep-1}$, then we can set $B_\pebtimestep = B_{\pebtimestep-1} \setminus \{ v \}$ and $W_\pebtimestep = W_{\pebtimestep-1}$.
	\item[White pebble placement on $v$:] A white pebble may be placed on any empty vertex at any time. Formally, if $v \not \in B_{\pebtimestep-1} \cup W_{\pebtimestep-1}$, then we can set $B_\pebtimestep = B_{\pebtimestep-1}$ and $W_\pebtimestep = W_{\pebtimestep-1} \cup \{ v \}$.
	\item[White pebble removal from $v$:] If all direct predecessors of a white-pebbled vertex $v$ have pebbles on them (black or white), the white pebble on $v$ may be removed. Formally, letting $B_\pebtimestep = B_{\pebtimestep-1}$ and $W_\pebtimestep = W_{\pebtimestep-1} \setminus \{ v \}$ is allowed if $v \in W_{\pebtimestep-1}$ and $\pred_G(v) \subseteq B_{\pebtimestep-1} \cup W_{\pebtimestep-1}$. In particular, a white pebble can always be removed from a source vertex.
\end{description}
A \emph{black-white pebbling} of $G$ is a sequence of pebble configurations $\Pebbling = (\PebConfig_0,\PebConfig_1,\dots,\PebConfig_\pebendtime)$ such that $\PebConfig_0 = \PebConfig_\pebendtime = 
(\emptyset,\emptyset),$ for some $i\leq \pebendtime,$ $z\in B_i\cup W_i$, and for all $\pebtimestep \in [\pebendtime]$ it holds that $\PebConfig_\pebtimestep$ can be obtained from $\PebConfig_{\pebtimestep-1}$ by applying exactly one of the above-stated rules.

A \emph{black pebbling} is a pebbling where $W_\pebtimestep = \varnothing$ for all $\pebtimestep \in [\pebendtime]$.
Observe that w.l.o.g. in a black pebbling we can always assume that $B_{\pebendtime-1}=\{z\}$. 
For convenience we will also use the dual notion of  \emph{white pebbling} game.
A {white (only) pebbling} is a pebbling where $B_\pebtimestep = \varnothing$ for all $\pebtimestep \in [\pebendtime]$. 
Notice that  $\Pebbling = (\PebConfig_0,\PebConfig_1,\dots,\PebConfig_\pebendtime)$ is a black pebbling of $G$ if and only if 
$\Pebbling' = (\PebConfig'_t,\dots,\PebConfig'_0)$ is a white
pebbling of $G$, where  each configuration $\PebConfig'_i$ contains the same 
set of pebbled vertices as in $\PebConfig_i$, but with white pebbles instead 
of black pebbles. In a white pebbling we can always suppose that $W_1=\{z\}.$
\end{defi}

\begin{defi}[Pebbling time, space, and price]
	The \emph{time} of a pebbling $\Pebbling = (\PebConfig_0,\PebConfig_1,\dots,\PebConfig_\pebendtime)$ is $\pebtime(\Pebbling) := \pebendtime$ and its \emph{space} is $\pebspace(\Pebbling) := \max_{\pebtimestep \in \nat{\pebendtime}} | B_\pebtimestep \cup W_\pebtimestep |$.
	The \emph{black-white pebbling price} (also known as the \emph{pebbling measure} or \emph{pebbling number}) of~$G$, which we will denote by~$\BW(G)$, is the minimum space of any black-white pebbling of~$G$.
	The \emph{black pebbling price} of~$G$, denoted by $\Black(G)$, is the minimum space of any black pebbling of~$G$. By the observation above, 
 the white pebbling price $\White(G)$ coincides with $\Black(G)$
\end{defi}

Finally, we mention the reversible pebble game introduced in~\cite{Bennett89TimeSpaceReversible}. In the reversible pebble game only black pebbles are placed and  the moves performed in reverse order should also constitute a legal black pebbling, which means that the rules for pebble placements and removals have to become symmetric. 

\begin{defi}[Reversible pebble game]
   The \emph{reversible pebble game} on~$G$ is the following version of  the game: At any time $\pebtimestep$ of the game, we have a pebble configuration $\PebConfig_\pebtimestep \subseteq V$. A configuration $\PebConfig_{\pebtimestep-1}$ can be changed to $\PebConfig_\pebtimestep$ by applying exactly one of the following rules:
   \begin{description}
        \item[Pebble placement on $v$:] If all direct predecessors of an empty vertex $v$ have pebbles on them, a pebble may be placed on $v$. $\PebConfig_\pebtimestep = \PebConfig_{\pebtimestep-1} \cup \{ v \}$ is allowed if $v \not \in \PebConfig_{\pebtimestep-1}$ and $\pred_G(v) \subseteq \PebConfig_{\pebtimestep-1}$. In particular, a pebble can always be placed on an empty source vertex $s$, since $\pred_G(s) = \varnothing$.
        \item[Reversible pebble removal from $v$:] If all direct predecessors of a pebbled vertex $v$ have pebbles on them, the pebble on $v$ may be removed. Formally, letting $\PebConfig_\pebtimestep = \PebConfig_{\pebtimestep-1} \setminus \{ v \}$ is allowed if $v \in \PebConfig_{\pebtimestep-1}$ and $\pred_G(v) \subseteq \PebConfig_{\pebtimestep-1}$. In particular, a pebble can always be removed from a source vertex.
    \end{description}
    
    A \emph{reversible pebbling} of $G$ is a sequence of pebble configurations $\Pebbling = (\PebConfig_0,\PebConfig_1,\dots,\PebConfig_\pebendtime)$ such that $\PebConfig_0 = \PebConfig_\pebendtime=\varnothing$, and $z\in \PebConfig_i$, for some $i<t$ and for all $\pebtimestep \in [\pebendtime]$ it holds that $\PebConfig_\pebtimestep$ can be obtained from $\PebConfig_{\pebtimestep-1}$ by applying exactly one of the above-stated rules.
 
\end{defi}

The notions of 
reversible pebbling time, space, and price are defined as in the other pebbling variants.   
From the definitions follows that for any DAG $G$ with a single sink,
$$\BW(G)\leq \Black(G)\leq \Rev(G).$$

\subsection{Formulas and Polynomials}

We will only consider propositional formulas in conjunctive normal form  (CNF). Such a  formula is a conjunction of clauses 
and a clause is a disjunction  of literals. A literal is a variable or its negation. For a formula $F$, $\var(F)$ denotes the set of its variables.

A Horn formula in a special type of CNF formula in which each clause has at most one positive literal.
For a more detailed treatment of formulas as well as the well known Resolution proof system we refer the 
interested reader to some of the introductory texts in the area like \cite{ST13SATProblem}.
We will basically only deal with pebbling formulas. These provide the connection between pebbling games and
proof complexity.

\begin{defi}[Pebbling formulas]
Let $G=(V,E)$ be a DAG with a set of sources $S \subseteq V$ and a unique sink $z$. We identify every vertex $v \in V$ with a Boolean variable $x_v$. For a vertex $v\in V$ we denote by  $\pred(v)$ the set of its direct predecessors. In particular, for a source vertex $v$, $\pred(v)=\emptyset$. The \emph{pebbling contradiction} over $G$, denoted $\peb(G)$, is the conjunction of the following clauses:
	\begin{itemize}
		\item for all  vertices $v$, the clause 
  $\bigvee_{u \in \pred(v)} \bar {x}_u \lor x_v$, 
         \hfill (\emph{pebbling axioms})
		\item 
  for the unique sink $z$, the unit clause $\bar {x}_z$. \hfill (\emph{sink axiom})
	\end{itemize}
\end{defi}

Observe that every clause in a pebbling formulas has at most one positive literal. These formulas are therefore Horn formulas.




A way to prove that a CNF formula is unsatisfiable is by translating it into a set of polynomials over a field   $\mathbb F$
 and then show 
that these polynomials do not have any common $\{0,1\}$-valued root. A clause 
$C=\bigvee_{x\in P} x \vee \bigvee_{y\in N} \bar y$ can be  encoded as the polynomial 
$p(C)=\prod_{x\in P} (1-x)  \prod_{y\in N} y$. A set of clauses $C_1,\dots,C_m$ is translated as
set of polynomials $p(C_1),\dots,p(C_m)$. Adding the polynomials $x_i^2-x_i$ (as axioms) for each variable $x_i$, there is no
common $\{0,1\}$-valued root for all these polynomials if and only if the original set of clauses is unsatisfiable. 
In this context we will consider a monomial $m$ as a set of variables and a polynomial $p$ as a linear combination of monomials. 
A monomial with its coefficient in $\mathbb F$
is called a monomial term.

When encoding the pebbling formulas as polynomials,
for a set $U\subseteq V$, we denote by  $m_U$ the monomial $\prod_{u\in U}x_u$. For $U=\emptyset,$ $m_U=1$. 
For every vertex $v\in V$ the axiom
$\bigvee_{u \in \pred(v)} \bar {x}_u \lor x_v$ becomes the polynomial $A_v:=m_{\pred(v)}(1-x_v)$, and  the sink
axiom $\bar {x}_z$ is transformed into the polynomial $A_{sink}:=x_z.$
Observe that every polynomial in the encoding of a pebbling formula is a monomial or the linear combination of two monomials. This is true for the polynomial encoding of every Horn formula.
To avoid confusion we  will denote the polynomial encoding of a CNF formula $F$ by $P_F.$ 


\subsection{Algebraic Proof Systems}

Several proof systems that work with polynomials  have been defined in the literature.  The simplest one is {\em Nullstellensatz}, \NS.
\begin{defi}

A Nullstellensatz refutation of the set of polynomials $p_1,\dots,p_m$ in $\mathbb F[x_1,\ldots, x_n]$ consists of a set of polynomials  
$g_1, . . . , g_m, h_1,\ldots, h_n $ such that
$$
\sum_{j=1,\ldots, m}p_jg_j+\sum_{i=1,\ldots, n}h_i(x_i^2-x_i)=1.
$$  

\end{defi}
As a consequence of Hilbert's Nullstellensatz, the  $\NS$ proof systems is sound and complete for the set of encodings of  unsatisfiable CNF formulas.

A stronger algebraic refutational calculus also dealing with polynomials  is the Polynomial Calculus ($\PC$).
As in the case of  Nullstellensatz, $\PC$ is intended to prove the  unsolvability of a set of polynomial equations.

\begin{defi}
The $\PC$ proof system uses  the following rules: 
\begin{description}
\item [\em Linear combination]
$$
\frac{p \qquad q}{\alpha p+ \beta q} \qquad \alpha,\beta \in \mathbb F. 
$$
\item [\em Multiplication]
$$
\frac{\quad p \quad}{x_ip} \qquad i \in [n]. 
$$
\end{description}
A refutation in $\PC$ of an initial unsolvable set of polynomials ${\mathcal P}$ is  
a sequence of polynomials $\{q_1,\dots, q_m\}$ such that each $q_i$ is either a polynomial in ${\mathcal P}$, a Boolean axiom ${x^2_i -x_i}$ or it is obtained by previous polynomials in the sequence applying one of the rules of the calculus. 
\end{defi}

\smallskip

A third less known  algebraic proof system  between $\NS$ and $\PC$ is Monomial Calculus, $\MC.$ This system 
was introduced in \cite{BerkholzG15}  identifying   exactly the complexity of refuting graph isomorphism formulas.
This proof system is defined like $\PC$ but the multiplication rule is only allowed to be applied to a monomial,
or to a monomial times an axiom.

\begin{defi}
The $\MC$ proof system uses  the following rules: 
\begin{description}
\item [\em Linear combination]
$$
\frac{p \qquad q}{\alpha p+ \beta q} \qquad \alpha,\beta \in \mathbb F. 
$$
\item [\em Multiplication]
    
$$
\frac{\quad p \quad}{x_ip} \qquad i \in [n],\ \ p\ \hbox{ is a monomial or the product of a monomial and an axiom.}
$$
\end{description}
As is the case of $\PC,$ a refutation in $\MC$ of an initial unsolvable set of polynomials ${\mathcal P}$ is  a sequence of polynomials $\{q_1,\dots, q_m\}$ where each one of them is either in ${\mathcal P}$,  an axiom or is obtained by   applying one of the rules of the calculus. 
\end{defi}

As  pointed out in \cite{BerkholzG15}, an equivalent definition of the  Nullstellensatz system, but a dynamic one,  
would be to restrict
the multiplication rule in the above definition even more, and only allow to apply it to polynomials that
are a  monomial multiplied by an axiom. In this way, the difference in the definition of the three systems NS, MC and PC is just a variation on how the multiplication rule can be applied:
multiplying a variable $x_i$ by any polynomial in the PC case, by  a monomial or by a monomial times an axiom in the MC case
and only by a monomial times an axiom in the NS case.
This alternative view of the definition also allows to consider configurational proofs in the $\NS$ system.

\begin{defi}[Configurational proofs]
Let ${\mathcal C}$ be one of the mentioned systems ${\mathcal C}\in\{\NS,\MC,\PC\}$.
For the space measures we need to define configurational proofs. Such a proof $\pi$ of a set of polynomials ${\mathcal P}$ in the system ${\mathcal C}$
is a sequence of configurations $\pi=C_0,\dots C_t$ in which each $C_i$ is a set of polynomials with $C_0=\emptyset$ and  $C_t=\{1\}$. Each configuration $C_i$ represents a set of polynomials that are kept simultaneously 
in memory at time $i$ in the refutation, and for each $i, 
0<i\leq t,$ $C_i$ is either

$C_{i-1}\cup\{p\}$ for some axiom $p$ (axiom download),  
 
$C_{i-1}\setminus \{p\}$  (erasure) or 

$C_{i-1}\cup\{p\}$ for some $p$ inferred from $C_{i-1}$ and ${\mathcal P}$
 by some rule of the system ${\mathcal C}$ (inference). 
\end{defi}

In order to analyze and compare refutations we will consider several complexity measures on them.

\begin{defi}[Complexity measures]
Let ${\mathcal C}$ be one of the mentioned systems ${\mathcal C}\in\{\NS,\MC,\PC\}$ and
let ${\pi}=\{q_1,\dots, q_m\}$ be a ${\mathcal C}$  refutation.
The degree of a  polynomial $q_i$, $\deg(q_i),$ is the maximum degree of its monomials and  
the \emph{degree} of ${\pi}$,  
$\deg_{\mathcal C}(\pi)=   
\max_{i=1,\ldots,m}(\deg(q_i))$. The size of ${\pi}$, denoted by 
$\Size_{\mathcal C}({\pi})$ 
is the total number of monomials in $\pi$ (counted with repetitions), when all polynomials $p_i$ are fully expanded as linear combinations of monomials\footnote{Usually the size in the $\NS$ proof system is defined
in a different way, for simplicity we keep this unifying definition although
in some of the referenced results the size of $\NS$ refutations corresponds
to the size definition given in the reference.}.

The variable space of the  proof $\pi$,  $\VS_{\mathcal C}({\pi})$  is defined as the maximum number of different variables appearing in any configuration of the proof.

For any of the defined complexity measures Comp and proof systems ${\mathcal C}$, and for every unsatisfiable set of polynomials 
$P_F$ we denote by Comp$_{\mathcal C}(P_F\vdash)$ the minimum over all ${\mathcal C}$ refutations of $P_F$ of
Comp$_{\mathcal C}(\pi).$
 \end{defi}

It is often convenient to consider a multilinear setting in which the multiplications in the mentioned algebraic systems are implicitly multilinearized.  Clearly the degree and size measures can only decrease in
this setting.

\section{Monomial Calculus and pebbling formulas}\label{MC}

In \cite{RezendeMNR21} it was shown that for any DAG $G$ with a single sink, the reversible pebbling  space 
and time of $G$, exactly coincides with the degree and the size of a NS refutation of $\peb_G$. We show that a very similar relation holds 
for the case of black pebbling and 
Monomial Calculus.

\begin{thm}\label{thm:peb-MC}
    Let  $G$  be a directed acyclic graph with a single sink $z$.
    If there is a black pebbling strategy of $G$ with time $t$ and space $s$
    then  there is a MC refutation of $\peb_{G}$ with  
    degree $s$ and size $ts$. The variable space of this refutation coincides with its degree.
\end{thm}

\begin{proof}
It is convenient to consider here the equivalent notion of white pebbling.
Let $\Pebbling = (\PebConfig_0,\dots,\PebConfig_\pebendtime)$ be a
white pebbling strategy for $G$ with $\PebConfig_1=\{z\}$ and 
$\PebConfig_t=\emptyset$ using $s$ pebbles. We show that for
each pebbling configuration $\PebConfig_i$, $i\in [t]$,  $\PebConfig_i=\{v_{i_1},\dots,v_{i_{k_i}} \}$ 
the monomial
$m_i=\prod_{v\in \PebConfig_i}x_{v}$ can be derived from $\peb_{G}$ and
$m_{i-1}$ in degree $s$ and size
 1 if $\PebConfig_i$ adds a pebble, or size $2s-1$ if 
 $\PebConfig_i$ removes a pebble.
This proves the result since in the $t$ steps of the 
pebbling strategy half of the steps add a pebble and the other half of
the steps remove a pebble (each added pebble has to be removed).
The total number of steps is therefore
$\frac{t}{2}+\frac{t}{2}(2s-1)=ts.$


    
{\bf Pebble placement:}
If the configuration at pebbling step $i+1$ is reached
after placing a white pebble on vertex $v$ and $\PebConfig_i=
\{u_{i_1},\dots,u_{i_{k_i}} \}$ with $k_i\leq s-1$ then 
$\PebConfig_{i+1}=\{v, u_{i_1},\dots,u_{i_{k_i}} \}.$ Multiplying the monomial  $m_i=\prod_{u\in \PebConfig_i}x_{u}$  by the variable
$x_v$ we obtain $m_{i+1}$. We have just added one more monomial of
degree at most $s$ to the proof.

{\bf Pebble removal:}
If the configuration at pebbling step $i+1$ is reached
after removing a white pebble from vertex $v$ and $\PebConfig_i=
\{v,u_{i_1},\dots,u_{i_{k_i}} \}$ with $k_i\leq s-1$ then all
predecessors  of  
$v$ are in the set $\{u_{i_1},\dots,u_{i_{k_i}} \}.$ 
For the derivation of $m_{i+1}$ we can multiply the axiom
$(1-x_v)\prod_{u\in\pred(v)}x_u$ by the variables in $\var(m_i)\setminus(\bigcup_{u\in\pred(v)} x_u\cup\{x_v\})$, 
and add this polynomial to $m_i$
obtaining $m_{i+1}$. Since there are at most $s-1$ variables in 
$\var(m_i)\setminus(\bigcup_{u\in\pred(v)} x_u\cup\{x_v\})$, 
the number of intermediate monomials added to the proof 
(counting also monomial $m_{i+1}$) is at most $2(s-1)+1=2s-1.$ 

Observe that in all the steps in the refutation, at most two different monomials appear and the number of different variables in these monomials coincides
with the largest of their degrees. This shows that the variable space of the 
MC refutation is bounded by $|\var(m_1)\cup\var(m_2)|\leq \max\{\deg(m_1),\deg(m_2)\} $.
\end{proof}

\begin{obs}
    The size bound $ts$ in the above proof comes from the way the MC rules
    are defined. As is the case of PC, in the multiplication rule only one variable at at time is multiplied, even when multiplying the axiom polynomials.
    When an axiom is multiplied by a monomial with several variables, all the
    intermediate polynomials contribute to the size of the MC refutation. 
    This is different from  the usual way to measure the size in the  NS case, where intermediate monomials are
    not counted. If we would define the MC rules as those in NS, that is, if a whole monomial could be multiplied by an axiom in
    one step, the size of the MC proof
    would avoid the  $s$ 
    factor in the monomial size and obtain size $\frac{3t}{2}$ instead. 
\end{obs}

In order to prove a result in the other direction 
we consider a very restricted kind of refutation in MC, analogous to what is known as an input refutation in Resolution. In this kind of refutation in every Resolution step one of the parent clauses must be an axiom. Input Resolution
is not complete, but it is complete for Horn formulas. We will show that the same is true for MC input refutations.

\begin{defi}
    A MC refutation $\pi$ of a contradictory set of polynomials $F$ 
    is called an {\emph {input  refutation}} if  there is a sequence of monomials 
     $M_0,\dots, M_t$ such that $M_0$ 
     is the product of a monomial and  an axiom, $M_t=1$ and for each $i$, $M_{i}$ is obtained by multiplying 
     $M_{i-1}$ with
     a variable, or by the linear combination rule from  $M_{i-1}$ and a monomial multiplied by an axiom polynomial.
     We will call the sequence of monomials $M_0,\dots, M_t$
    the backbone of the proof.
    \end{defi}

\begin{lem}\label{lemma-input-mc-proof}
    Let $F$ be an unsatisfiable  Horn formula and let $P_F$ be the encoding of $F$ as a  
    set of polynomials.   
    Let $\pi$ be any 
    MC refutation of $P_F$. There is an input   MC 
    refutation $\pi'$ of $P_F$ with at most the same size and   degree as $\pi.$  
 \end{lem}

\begin{proof}
Let $d$ and $t$ be the degree and size of $\pi$.  We can suppose that $\pi$ is multilinear. 
    We prove the result by induction on $k$, the number of times  the multiplication rule is applied to a monomial derived in $\pi$. In the base case $k=0$, $\pi$ is just a NS refutation of $P_F.$ 
    This means that there is a linear combination of a set of polynomials $S$ 
    that adds up to 1. Each of these polynomials 
    has the form of a polynomial axiom multiplied by a monomial and since $F$ is a Horn formula, each polynomial in $S$ has either one or two monomials with coefficients in $\{1,-1\}$. 
    We will represent such a  polynomial $p=\alpha_m m+\alpha_{m'}m'$ by the 
pair of monomials $(m,m')$. 
Since we are representing monomials by their set of variables,
the monomial $1$ is represented by $\emptyset$.
Clauses without positive literals are encoded as single monomials $(m)$.
Some  polynomial in $S$ has a single monomial  otherwise the whole set $S$ 
would have a common root by setting all variables to 1. 
Moreover, 
we claim there has to be a sequence of polynomials  $p_0,\dots,p_\ell$ in $S$ represented by the  monomials $(\emptyset ,m_1),(m_1,m_2),(m_2,m_3) \dots, (m_{\ell-1},m),(m).$ 
In order to see this, we define an undirected graph whose vertices are the different monomials in $S$ and there is an edge between two monomials $m,m'$ if both appear in the same polynomial in $S$. In a minimal $\MC$ proof we only need to consider the connected component containing monomial $1$. Such component must have a vertex $v$ coming
from a polynomial with a single monomial $m$ in $S$ because any set of binomials in $S$ can be satisfied by assigning all variables to 1. Therefore there is a path in the graph connecting vertices $1$ and $v$. Such a path with minimal length 
defines a sequence of edges $(1,v_1),(v_1,v_2),\dots, (v_{\ell-1},v)$, and 
from it we can
build the chain of polynomials $(\emptyset ,m_1),(m_1,m_2),(m_2,m_3) \dots, (m_{\ell-1},m),(m)$. This proves the claim.

Now we can define the input monomial refutation $\pi'$ starting at $M_0=m$  and applying then $\ell$ linear combinations with axioms multiplied by monomials and deriving all the monomials
$m_\ell,\dots,m_1$ until $1$ is derived. 
Observe that the monomials $M_0,\dots M_t$ are exactly those appearing 
in $p_0,\dots,p_{\ell}$. By the minimality of the sequence we also know that the monomials in
the backbone are all different since if $M_i=M_j$ for $i<j$,  we could shorten $\pi'$ by connecting $M_i$ with $M_{j+1}$.

All the monomials in $\pi'$ belong also to $\pi$, therefore 
the degree of the new refutation is not larger than that in $\pi.$
In fact all the polynomials in $p_0,\dots,p_{\ell}$ are already
in $\pi$. Besides these polynomials $\pi'$ contains also the
$\ell$ new monomials in the backbone. Since the polynomials $p_0,\dots,p_{\ell}$  belong to $\pi$ and in each linear combination of two polynomials at  most one monomial vanishes,
there are at least $\ell$ intermediate polynomials in $\pi$ until
1 is reached. This means that the size of $\pi'$ is bounded by $t.$

For the case $k>0$  let $m'$ be the first monomial in the proof
that is the result of a multiplication from a derived monomial 
$m$ and a variable $x$,  
 $m'=xm$ in $\pi$. The same 
argument as above shows that there is a sequence of polynomials $p_0,\dots,p_{\ell},\hat{m}$ in $\pi$ from which we can extract 
an input monomial refutation that starts at $M_0=\hat{m}$
and derives at some point $M_i=m$. 
In the next step the multiplication rule is applied to
obtain $M_{i+1}=m'.$ Observe that the set of polynomials  
$m'\cup P_F$ still has the Horn property and that there is a 
sub-proof of $\pi$ that derives the monomial 1 from this set 
applying 
the multiplication rule at most $k-1$ times.   
By induction hypothesis  we know that there
is a sequence of polynomials $p'_0,\dots,p'_{\ell'}$ in $\pi$ represented by the  monomials $(\emptyset ,m'_1),(m'_1,m'_2), \dots, (m'_r,m')$ from which 
an input refutation of $P_F\cup m'$ can be extracted. 
We can put together both input MC refutations $M_0\dots M_i$
and $M_{i+1},\dots,1$. Again we can assume that all the monomials in the backbone are different since
otherwise we could shorten $\pi'$. 
By the same argument as in the base case the size and degree
of the input MC refutation cannot be larger than that of $\pi.$
\end{proof}

Since pebbling formulas are Horn formulas we immediately obtain:

\begin{cor}\label{cor:input-mc-proof-of-peb}
    Let $G$ be a directed acyclic graph with a single sink vertex $z$ and 
    let $\pi$ be a MC refutation of $\peb(G).$ There is an input   MC 
    refutation $\pi'$ of $\peb(G)$ with at most the same size and   degree as $\pi.$  
 \end{cor}

We show next that from an MC refutation of a pebbling formula it is possible to extract a black pebbling strategy for the corresponding graph.

\begin{thm}\label{thm:MC-ref-peb}
    Let $G$ be a directed acyclic graph with a single sink. Let $\pi$ be a MC refutation of
    $\peb(G)$ with degree
    $s$ and size $t$. There is a 
    black pebbling strategy for $G$ with  $s$ pebbles and time $t+s$.
\end{thm}

\begin{proof}
Because of Corollary~\ref{cor:input-mc-proof-of-peb} we can suppose that there exists  an input MC refutation
with monomials $M_0,\dots M_t$ starting with $M_1=mx_{\rm sink}$
for some monomial $m$ and with $M_t=1$. We describe a strategy for a white pebbling of $G$ following $\pi$.
As explained in the preliminaries, this is equivalent to a black pebbling strategy.
At each step $i$ only the vertices corresponding to variables in $M_i$ have a pebble on them.
In a multiplication step a new pebble is added, which is always possible in a white pebbling strategy. We
only have to show that in case variables disappear when going from $M_i$ to $M_{i+1}$, this is also a correct pebbling move. But in this case, the step from $i$ to $i+1$ is a linear combination of  $M_i$ with the axiom for
some vertex  $v$,
$m_{\pred(v)}(1-x_v)$ multiplied by some monomial $m$. The only variable that can disappear in $M_{i+1}$ is
$x_v$ and in this case the monomial $M_i=m_{\pred(v)}x_v$. Therefore all the vertices in $\pred(v)$ have pebbles on them
and the pebble in $x_v$ can be  removed.
At the end of the refutation, when the 1 monomial is reached there are no pebbles left on $G.$
The number of pebbles present at any moment is the number of variables  in any of the monomials and
this is the degree of $\pi$. The number of pebbling steps needed is 
at most $d$ steps to place  a pebble in each variable of 
$M_1=mx_{\rm sink}$ and then $t$ more pebbling steps.
\end{proof}

\begin{obs}
    For the case of Polynomial Calculus it is known that strong degree lower bounds imply size lower bounds. If 
    a set of unsatisfiable polynomials $P_F$ with $n$ variables and constant degree requires 
     PC refutations of degree $d$, then any PC refutation
    of $P_F$ requires size at least $2^{\Omega(\frac{d^2}{n})}$
    \cite{IPS99LowerBounds}.
    The previous results show that this does not hold for Monomial Calculus. This follows from the fact that there are graph families $\{G_n\}_{n=0}^\infty$ with $n$ vertices and constant in-degree that require black pebbling space 
    $\Omega(\frac{n}{\log n})$ \cite{PTC76SpaceBounds}.
    Theorem~\ref{thm:MC-ref-peb} implies that the pebbling formulas
    for this graph family needs degree $\Omega(\frac{n}{\log n})$.
    On the other hand, for every single-sink DAG with $n$ vertices there is a trivial
     black pebbling strategy using space  $n$ and  pebbling time $2n$.
    By Theorem~\ref{thm:peb-MC} this implies that 
    the pebbling formulas corresponding to the 
    graphs in $\{G_n\}_{n=0}^\infty$ have MC refutations of quadratic size in $n$. 
    This is a family of formulas with MC refutation degree $\Omega(\frac{n}{\log n})$ but having quadratic size refutations, a very different situation from the one in the PC case.
\end{obs}

\subsection{Degree separations}\label{subsec:degree-sep}

The given relationships between MC and the black pebbling game allow for 
the immediate translation of pebbling results to Monomial Calculus. We start
with some degree separations between MC and PC. 
The original motivation for introducing   MC was the close connection between
the degree complexity of the refutation of the graph isomorphism formulas in this proof system, 
and the Weisfeiler-Leman hierarchy
\cite{BerkholzG15}.  Formulas corresponding to  non-isomorphic graphs pairs that can
only be distinguished using 
a large level of the WL algorithm, require a MC refutation with large degree. 
It  was proven
later in \cite{AF23,GGPP19}, that the degree of a PC refutation of the isomorphism formulas
cannot be much smaller than in the MC case,  in fact the degrees of a MC and a PC refutation can only be
a constant factor apart. We improve this separation and obtain 
an almost optimal degree separations by considering 
the pebbling formulas.
In \cite{BCIP02Homogenization} it was shown
that pebbling formulas have constant PC degree and that  
for any directed acyclic graph $G$ with black pebbling price $B(G),$ the formula 
$\peb(G)$ 
requires  NS refutations with degree $\Omega(B(G))$.  Since it is known that
there are graph families $\{G_n\}_{n=0}^\infty$ with $\Theta(n)$ vertices and 
$B(G_n)=\Omega(\frac{n}{\log n})$ \cite{PTC76SpaceBounds}, this implies a
degree separation of $\Omega(\frac{n}{\log n})$ between PC and NS. From
Theorem~\ref{thm:MC-ref-peb} follows that this is in fact a degree separation between MC and PC.

\begin{thm}
There is an unsatisfiable 
family of formulas $\{F_n\}_{n=0}^\infty$ with $\Theta(n)$ variables each, that have PC refutations
of constant degree but require MC refutations of degree 
$\Omega(\frac{n}{\log n})$.
\end{thm}

For the case of NS, from Theorem~\ref{thm:peb-MC} and the equivalence  between reversible pebbling price and NS degree from \cite{dRMNPRV20}, \cite{RezendeMNR21}, follows that a separation
between reversible and black pebbling price for a graph family implies a
degree separation between  NS and MC for the corresponding pebbling formulas.
For example it is known that a directed  path graphs with $n$ vertices can be black pebbled with 2 pebbles but requires reversible pebbling number 
$\lceil \log n\rceil$ \cite{Bennett89TimeSpaceReversible}. Translated to pebbling
formulas this means:

\begin{thm}\label{degree-2}
    There is a
family of unsatisfiable formulas $\{F_n\}_{n=0}^\infty$ with $\Theta(n)$ variables each, that have MC refutations
of  degree 2 but require NS refutations of degree 
$\lceil \log n\rceil$.
\end{thm}

There are other
graph families for which a separation between the black and reversible pebbling prices by a logarithmic factor in the number of vertices is known,
\cite{CLNV15Hardness},\cite{ToranW21}. The separations in pebbling for these graphs is translated into the next result.

\begin{thm}
    For any function $s(n)=O(n^{1/2-\epsilon})$ for constant $0<\epsilon<\frac{1}{2}$ there is a
family of unsatisfiable formulas $\{F_n\}_{n=0}^\infty$ with $\Theta(n)$ variables each, that have MC refutations
of  degree $O(s(n))$ but require NS refutations of degree  
$\Omega( s(n) \log n)$.
\end{thm}

The question of whether the separation between 
reversible and black pebbling space can be larger than a logarithmic factor in
the number of nodes is open.
The best known degree separation between  NS and and MC is slightly better. This was obtained in \cite{GroheP17} with very different methods. Using a classic result from descriptive complexity \cite{Immerman81}, the authors show that for every constant
$c\geq 1$ there are families of formulas $F_n$ with $O(n)$
 variables that have a degree 3 MC refutation but require NS degree at least
 $\log^c({n}).$ It is also open whether this degree separation between NS
and MC is optimal. Theorem~\ref{degree-2} is the best known degree separation between NS and an MC refutation of degree 2.

\subsection{Size-degree tradeoffs for MC}

The close connections between black pebbling space and monomial calculus 
expressed in Theorems~\ref{thm:peb-MC} and \ref{thm:MC-ref-peb} make it possible
to translate space-time tradeoffs for pebbling into degree-size tradeoffs for MC. 
There is a slight loss of the time parameter that comes from the extra space factor in the MC refutation from Theorem~\ref{thm:peb-MC}.
We present several such results as examples. The first one is an extreme tradeoff result that shows how decreasing the MC degree by one in a refutation can make the size increase
exponentially.

\begin{thmC}[\rm{\cite{Savage98}}]
There is a family of directed 
graphs $\{G_n\}_{n=0}^\infty$ having $\Theta(n^2)$ vertices each and with
$\Black(G_n)=\Theta(n)$ for which any black pebbling strategy with 
$\Black(G_n)$ pebbles
requires at least $2^{\Omega(n\log n)}$ steps while there is a pebbling strategy
with $\Black(G_n)+1$ pebbles and $O(n^2)$ steps. 
\end{thmC}

\begin{cor}
There is  a family of unsatisfiable formulas  $\{F_n\}_{n=0}^\infty$ with $F_n$ having $O(n^2)$ variables and 
 $d_n\in O(n)$ such that $F_n$ has a MC refutation of degree $d_n$ 
 but any MC refutation with this degree requires size $2^{\Omega(n\log n)}.$
 On the other side there is  a MC refutation of $F_n$ with degree $d_n+1$ and size O$(n^3).$ 
\end{cor}

For  a second example we use  a robust time-space result from \cite{Nordstrom09PebblingSurvey}.

\begin{thm}
    There is a family of directed 
graphs $\{G_n\}_{n=0}^\infty$ having $\Theta(n)$ vertices each and with
$\Black(G_n)=O(\log^2 n),$ with a black pebbling strategy having simultaneously space $O(n/\log n)$
and time $O(n)$. There is  also a constant $c>0$ for which any
pebbling strategy using less than $cn/\log n$ pebbles requires at least $n^{\Omega(\log\log n)}$ steps. 
\end{thm}

\begin{cor}
There is  a family of unsatisfiable formulas  $\{F_n\}_{n=0}^\infty$ with $F_n$ having $O(n)$ variables, and a constant $c>0$ 
such that $F_n$ has a MC refutation of degree 
$O(n/\log n)$ and size $O(n^2/\log n)$ simultaneously 
 but for which  any MC refutation with  degree smaller than 
 $cn/\log n$  requires size at least $n^{\Omega(\log\log n)}$. 
 \end{cor}




We give a final example for a robust degree-size tradeoff for arbitrary small non-constant degree.

\begin{thmC}[\rm{\cite{Nordstrom12RelativeStrength}}]
    Let $g(n)$ be any arbitrarily slow growing function 
    $\omega(1)\leq g(n) =O(n^{1/7})$, and let $\epsilon >0$. There is a family of directed graphs 
    $\{G_n\}_{n=0}^\infty$ having $\Theta(n)$ vertices each and with $\Black(G_n)=2g(n)+O(1)$, for which there is a black pebbling strategy for $G_n$ in time $O(n)$ and space $O(n/g^2(n))^{1/3}$ simultaneously. Moreover,  any pebbling strategy for $G_n$ with space at most $O(n/g^2(n))^{1/3-\epsilon}$ 
    requires superpolynomial time in $n$.
\end{thmC}

\begin{cor}
    Let $g(n)$ be any arbitrarily slow growing function 
    $\omega(1)\leq g(n) =O(n^{1/7})$, and let $\epsilon >0$. There is a family of unsatisfiable formulas  
    $\{F_n\}_{n=0}^\infty$ with $F_n$ having $O(n)$ variables 
    that have MC refutations with degree $2g(n)+O(1)$. For these formulas   there are MC refutations
    of size  $O(n^2)$ and degree $O(n/g^2(n))^{1/3}$ simultaneously, but any MC refutation with
    degree at most $O(n/g^2(n))^{1/3-\epsilon}$ 
    requires superpolynomial size in $n$.
\end{cor}

\section{Pebble Games and Variable Space}

The equality between degree and pebbling price for the cases of 
Monomial Calculus and black pebbling from the previous section, as well as for Nullstellensatz and reversible pebbling from \cite{RezendeMNR21}
cannot be extended to the case of Polynomial Calculus and black-white
pebbling price since as already mentioned, it was proven in \cite{BCIP02Homogenization} that for any DAG $G$, $\deg_{\PC}(\peb(G))=O(1).$
We show in this section that the correspondence between the three pebbling
variations and the proof systems holds if we consider the variables space measure instead.

It can be seen in the proof of 
Theorem~\ref{thm:peb-MC}, that not only the 
minimum degree of a monomial calculus refutation of $\peb(G)$, but also the 
minimum variable space is bounded by  the black pebbling price of $G$. 
The same can be observed in  the proof of Theorem~3.1 in \cite{RezendeMNR21}
for the case of Nullstellensatz and reversible pebbling.
Considering the trivial fact that variable space measure  
is always greater or equal than the 
degree needed for the refutation of a formula in all three proof systems
$\NS,\MC$ and $\PC$, and considering Theorems~\ref{thm:MC-ref-peb} as well as Theorem~3.4
in \cite{RezendeMNR21} this implies:

\begin{obs}\label{obs:vs-peb}
For every DAG $G$ with a single sink, $\VS_{\NS}(\peb(G)\vdash)=\Rev(G)$ and
$\VS_{\MC}(\peb(G)\vdash)=\Black(G)$.
\end{obs}

For the case of black-white pebbling it is known that 
for the Resolution proof system, the variable space needed in a refutation
of $\peb(G)$ equals $\BW(G).$ The inclusion from left to right is from
\cite{Ben-Sasson02SizeSpaceTradeoffs} while the other inclusion 
appeared in \cite{AlexHertel08Thesis}. This results can be extended to 
stronger  proof systems using the following result:

\begin{lem}[\cite{BNT13SomeTradeoffs}, \cite{Razborov18SpaceDepth}]\label{lem:variable-space}
Let  $S$ be a proof system that  can simulate Resolution step
by step without including new variables.   
For every unsatisfiable formula
$F$,  $\VS_S(F\vdash\nolinebreak) = \VS_{\Res}(F\vdash)$. 
\end{lem}

This implies:

\begin{obs}
For every DAG $G$ with a single sink, $\VS_{\PC}(\peb(G)\vdash)=\BW(G)$. 
\end{obs}

Together with Observation~\ref{obs:vs-peb} this shows the equivalence between variable space  in the proof systems and the pebbling price in the 
three variations of the game.

\subsection{Variable Space Separations}

These observations allow us to use pebbling results to obtain separation
in the variable space complexity in the algebraic proof systems.
The reason why these results do not contradict Lemma~\ref{lem:variable-space}, is that $\MC$ (and $\NS)$ cannot simulate
Resolution step by step since the intermediate polynomials in the simulation are not necessarily monomials.  

For the variable space separations between $\NS$ and $\MC$ on pebbling formulas, the same 
degree separations given in Subsection~\ref{subsec:degree-sep}  hold, since
as we have seen, for this kind of formulas the variable space and the degree
coincide in both proof systems.
For the case of $\MC$ versus $\PC$, it is known that for any DAG $G$, the separation between the black and
the black-white pebbling prices can be at most quadratic \cite{MadH81ComparisonOfTwoVariationsOfPebbleGame}. This limits
the variable space gap  between  $\MC$ and $\PC$ that can be obtained 
using pebbling formulas. In \cite{Wilber88WhitePebblesHelp} a family of graphs is given that shows an asymptotic separation between the black-white
and black pebbling prices. Translating this to our context we obtain:

\begin{thm}
    There is  a family of unsatisfiable formulas $\{F_n\}_{n=0}^\infty$ with polynomially many variables (in $n$) such that $\VS_\PC(F_n\vdash)=O(n)$ and 
    $\VS_\MC(F_n\vdash)=\Omega(\frac{n\log n}{\log\log n}).$
\end{thm}

An optimal quadratic  separation between the black-white and black pebbling price was given in \cite{KS91OnThePowerOfWhitePebbles} but for a family of 
graphs having exponentially many vertices respect to their pebbling price. 
This implies:

\begin{thm}
    There is  a family of unsatisfiable formulas $\{F_n\}_{n=0}^\infty$ with 
    $\exp(\Theta(n\log n))$ many variables  such that $\VS_\PC(F_n\vdash)=O(n)$ and
    $\VS_\MC(F_n\vdash)=\Omega(n^2).$
\end{thm}

\section{Conclusions and Open Questions}

We have proven a strong connection between the black
pebble game and the Monomial Calculus proof system by showing that the degree and
size bounds required simultaneously in a MC  refutation of the pebbling formula
for a DAG $G$
 closely correspond to the number of pebbles and the time in a pebbling
strategy for $G$. Together with the connections between the complexity of MC refutations and the Graph Isomorphism
formulas from \cite{BerkholzG15}, these results underline the importance of 
MC as a natural proof system between NS and PC.
Our results  improve the known relations between 
the complexities of pebble games and algebraic proof systems and
implies strong degree-size tradeoffs for the $\MC$ system as well as
degree separations between $\NS, \MC$ and $\PC$.

We have also shown that the variable space measure for the refutation of
pebbling formulas in the three
systems $\PC, \MC$ and $\NS$ exactly corresponds to the number of pebbles in the black-white, black and reversible games. From this
equivalence we obtain variable space separations between the proof systems. 

It is open whether these separations are optimal or can be improved using other techniques. Finding out what is the optimal degree separation between the $\NS$ and $\MC$ proof systems is another
interesting open question.

\bibliography{references3.bib}
\bibliographystyle{alphaurl}

\end{document}